\documentclass{article}
\usepackage[utf8]{inputenc}
\usepackage{authblk}
\usepackage{setspace}
\usepackage[margin=1.25in]{geometry}
\usepackage{graphicx}
\graphicspath{{./Figures/}}
\usepackage{subcaption}
\usepackage{amsmath}
\usepackage{lineno}

\usepackage{ragged2e}
\usepackage{physics, bm, siunitx}
\usepackage[style=nejm, 
citestyle=numeric-comp,
sorting=none]{biblatex}
\title{Single Spatio-Temporal Mode Bright Twin-Beam Source Across the Near- and Mid-Infrared}

\author[1,*]{Gabriel Demontigny}
\author[1]{Patrick Cusson}
\author[1]{Amauri Perraton-Elorza}
\author[1,2]{Esteban Murillo-Zapata}
\author[1,3]{Eli Martel}
\author[4,5]{Andrei Rasputnyi} 
\author[4,5,6]{Maria Chekhova} 
\author[1,*]{St\'{e}phane Virally}
\author[1,7,*]{Denis Seletskiy}

\affil[1]{Laboratoire femtoQ, D\'{e}partement de G\'{e}nie Physique, Polytechnique Montr\'{e}al, 2500 Chemin de Polytechnique, Montr\'{e}al QC, H3T 1J4, Canada}
\affil[2]{Departamento de Física, Universidad Del Valle, Comuna 17, Cali, Valle del Cauca, Colombia}
\affil[3]{Institut Interdisciplinaire d'Innovation Technologique, Universit\'{e} de Sherbrooke, 3000 boulevard de l'Universit\'{e}, Sherbrooke, QC, Canada}
\affil[4]{Max Planck Institute for the Science of Light, Erlangen, Germany}
\affil[5]{Department of Physics, Friedrich-Alexander-Universit\"{a}t
Erlangen-N\"{u}rnberg (FAU), Erlangen, Germany.}
\affil[6]{Faculty of Electrical and Computer Engineering, Technion—Israel
Institute of Technology, Haifa 32000, Israel}
\affil[7]{femtoQ Laboratory, Department of Physics and Astronomy, University of New Mexico, 210 Yale Blvd. NE, Albuquerque, NM 87106, United States}

\affil[*]{gabriel.demontigny@polymtl.ca, stephane.virally@polymtl.ca, denisel@unm.edu}

\date{}
\begin{document}

\maketitle

\begin{abstract}
We introduce an ultrafast, bright, entangled twin-beam source generated by type-0 parametric down-conversion in periodically-poled lithium niobate at MHz repetition rate, with continuously tunable Schmidt number $K$ set by the pump pulse duration. Photon-number statistics characterization via $g^{(2)}(0)$ and singular-value decomposition of the signal spectral density matrix yield $K\simeq1.05$ and $K\simeq1.03$, respectively, maintained over multiple orders of magnitude in brightness. Group-delay dispersion of the pump drives a continuous transition from single-mode operation to a controlled multimode regime, consistent with the temporal gain window departing from the inverse phase-matching bandwidth.
Strong non-degeneracy of the source (signal at \SI{1.37}{\um}, idler at \SI{4.0}{\um}, $\sim\SI{100}{\fs}$ duration) decouples a mid-infrared interaction wavelength, which overlaps with molecular vibrational resonances, from a near-infrared detection band, establishing a practical platform for quantum-enhanced metrology, nonlinear interferometry, and mid-infrared spectroscopic sensing.
We show that in the bright few-mode limit, the total entanglement resource is clearly separated between \emph{modal} and \emph{occupational} degrees of freedom, and that our source allocates up to \SIrange{95}{97}{\percent} of that resource to the occupational sector.

\end{abstract}

\section{Introduction and Motivation}

Nonclassical light sources that are simultaneously \emph{bright}, \emph{ultrafast}, and \emph{mode-controlled} are of great interest for quantum-enhanced metrology, nonlinear interferometry, and spectroscopic sensing. In particular, non-degenerate parametric down-conversion (PDC) produces two-mode squeezed vacuum whose quantum correlations between signal and idler~\cite{kwiat_hyper-entangled_1997} are advantageous in many applications, including quantum-enhanced metrology and secure communications. In the high-gain regime of PDC, the physics is most often discussed in terms of \emph{bright squeezed vacuum} (BSV)~\cite{chekhova_bright_2015}, which corresponds to the degenerate case of PDC--- where signal and idler are indistinguishable in the detection setup---and \emph{bright twin beams} (BTB)~\cite{finger_2015} in the non-degenerate case---where the observables associated with signal and idler can be detected separately. Non-classical correlations persist even at those macroscopic photon numbers, enabling sub-shot-noise measurements when single-photon counting might be impractical~\cite{chekhova-PNDS,finger_2015}. Prior work has clarified how increasing the gain reshapes the spectral and correlation properties of PDC emission, including gain-dependent changes of bandwidth and pairwise spectral correlations~\cite{spasibko_spectral_2012,florez_pump_2020,sharapova_2020,quesada_beyond_2022}.
Near-single-mode operation ensures that the generated state is a single
entangled set of twin beams rather than a tensor product of independent entangled sets across orthogonal Schmidt modes~\cite{law_continuous_2000, Law2004PRL} (theory extended to bright twin beams in Supplementary Material Secs. S3, S4, S5).

As developed in Sec.~\ref{sec:Partitioning}, near-single-mode operation makes the entanglement resource fully accessible without mode-sorting by direct broadband detection in continuous-variable~\cite{braunstein_quantum_2005} and nonlinear-interferometric schemes~\cite{yurke_su2_1986,ou_quantum_2020,lemos_quantum_2014}, and in photon-number conditioning protocols such as electro-optic sampling (EOS)~\cite{virally_enhanced_2021} (see Supplementary Material Secs. S3, S4). In contrast, controlled multimode operation supports multiplexed quantum-information distribution across classical degrees of freedom such as frequency, time, or transverse profile.

Bright squeezed vacuum has also recently emerged as a \emph{driver} of strong-field and nonlinear processes, with the non-classical photon statistics of the driving field imprinted onto the strong-field response across platforms ranging from perturbative harmonic generation~\cite{spasibko_multiphoton_2017} and non-perturbative high-harmonic generation in solids~\cite{rasputnyi_hhg_2024} to quantum-controlled high-harmonic sideband emission~\cite{lemieux_bunching_2025} and multiphoton electron emission from nanostructures~\cite{heimerl_multiphoton_2024,heimerl_strongfield_2025}. This is supported by theoretical predictions of modified cutoff laws, photon-statistics forces, and squeezed harmonics~\cite{gorlach_hhg_2023,eventzur_force_2023,eventzur_squeezed_2024,eventzur_motion_2024}. Reciprocally, the high-harmonic process itself generates radiation carrying non-classical statistics including squeezing and entanglement~\cite{theidel_evidence_2024,stammer_entanglement_2024}, and BSV quantum correlations have been transferred onto XUV attosecond pulses, enabling attosecond-resolved measurements of quantum fluctuations in both light and matter~\cite{eventzur_attosecond_2025}. For both directions, single-mode operation maximizes the peak quantum field amplitude available to the strong-field interaction while preserving the non-classical statistics on which these processes depend.

There also exists a strong motivation for sources of \emph{strongly non-degenerate} twin beams because they decouple the wavelength best suited for \emph{interaction} from the wavelength best suited for \emph{detection}. This separation is particularly compelling when one twin reaches mid-infrared (mid-IR) wavelengths that overlap with molecular vibrational resonances. The absorbance of the mid-IR twin is imprinted onto the near-IR twin, which is generated to overlap with high-quantum-efficiency, low-noise detectors. This idea also underlies nondegenerate nonlinear interferometry, notably SU(1,1) interferometers \cite{yurke_su2_1986,ou_quantum_2020,santandrea_lossy_2023}, and ``undetected-photon'' sensing and imaging \cite{lemos_quantum_2014,kalashnikov_2016,chekhova_2016}. 

Here we demonstrate an ultrafast, bright, entangled, strongly non-degenerate twin-beam source in the near- and mid-infrared, generated by type-0 parametric down-conversion (PDC) in periodically-poled lithium niobate (PPLN) crystal. The signal is centered near \SI{1.37}{\um} and the idler near \SI{4.0}{\um}, with pulse durations on the order of \SI{100}{\fs}. We verify near-single spatio-temporal mode operation over multiple orders of magnitude in brightness using two independent diagnostics of the signal beam: photon-number statistics via the second-order correlation function at zero delay $g^{(2)}(0)$ and singular-value (or Schmidt) decomposition of the measured spectral intensity covariance. Finally, we identify the pump pulse duration as a key control parameter: increasing it via added group-delay dispersion (GDD) drives a transition from single-mode to multimode amplification, consistent with the intuitive requirement that the temporal gain window be matched to the inverse phase-matching bandwidth, most directly understood through the picture of space-time volumetric amplification of the input vacuum state \cite{riek_direct_2015,moskalenko_paraxial_2015}.

\section{Modes of Bright Twin-Beams}\label{sec:Modes}

Bright twin-beam states are generally multimode (in the sense of first quantization) in both space-wavevector and time-frequency, and the effective number of occupied modes can evolve with gain, pump parameters, and collection geometry~\cite{sharapova_2020, quesada_beyond_2022}. In the case of bright sources, a high number of modes can be deleterious in at least two respects: a reduction in squeezing for all modes, due to the exponential nature of SPDC gain and the limited amount of pump power; and a requirement for experimentally difficult and lossy mode sorting if one wants to recover and exploit all quantum correlations.
Among other things, this is why achieving one of the earliest record brightness of twin-beams (2500 photons per mode) required a fiber source with only 3 modes~\cite{finger_2015}.

A natural framework to quantify the mode structure is provided by the Schmidt decomposition of the joint spectral amplitude \cite{brecht_photon_2015,raymer_temporal_2020,fabre_modes_2020},
\begin{equation}
    \Psi(\bm{k}_\textrm{s}, \bm{k}_\textrm{i}) = \sum_n \sqrt{\lambda_n}\,\psi_n(\bm{k}_\textrm{s})\,\phi_n(\bm{k}_\textrm{i}),
    \label{eq:Psi}
\end{equation}
where the $\bm{k}_\textrm{s,i}$ are the wavevectors of the signal and idler, respectively. The Schmidt decomposition defines orthonormal broadband ladder operators $\hat{A}_n$ and $\hat{B}_n$ for the signal and idler---see Eq. S7---along with spatio-temporal mode structures
\begin{equation}
\begin{split}
    &\psi_n(\bm{r},t)=\int\frac{\dd^3\bm{k}}{\sqrt{8\pi^3}}\,\psi_n(\bm{k})\,e^{i\bm{k}\cdot\bm{r}-i\omega(\bm{k})t};\\ &\phi_n(\bm{r},t)=\int\frac{\dd^3\bm{k}}{\sqrt{8\pi^3}}\,\phi_n(\bm{k})\,e^{i\bm{k}\cdot\bm{r}-i\omega(\bm{k})t},
    \end{split}
    \label{eq:psiphit}
\end{equation}
where the pulsation $\omega(\bm{k})$ is imposed by the dispersion relation. The effective number of occupied spatio-temporal modes is commonly quantified via the Schmidt number
\begin{equation}
    K=\frac{1}{\sum_n \lambda_n^2},
\end{equation}
with $K \rightarrow 1$ indicating near-single-mode operation.

The joint spectral amplitude $\Psi(\bm{k}_s,\bm{k}_i)$ of \eqref{eq:Psi} provides the most complete characterization of the Schmidt-mode basis (with gain-dependent populations developed in Sec.~\ref{sec:Partitioning}). It can be obtained directly via the correlator $\ev{\hat{a}(\bm{k}_\textrm{s})\hat{a}(\bm{k}_\textrm{i})}$, whose measurement is challenged by the requirement of phase-sensitive detection of both beams.
It is however possible to trace over the idler beam and the spatial degrees of freedom to measure instead the reduced spectral density matrix of the signal beam, $G^{(1)}_\textrm{s}(\omega_\textrm{s},\omega_\textrm{s}') = \langle \hat{a}^\dagger(\omega_\textrm{s})\hat{a}(\omega_\textrm{s}')\rangle$. It is also diagonal in the Schmidt basis of signal modes (see Supplementary Material Sec. S3) and has the advantage of being experimentally accessible via shot-to-shot spectra on the signal beam. The spectral intensity profiles of the Schmidt modes of the signal beam are obtained via singular value decomposition of $G^{(1)}_\textrm{s}$. The bilinear structure of the parametric Hamiltonian ensures that this eigenbasis is gain-independent, although the eigenvalues are not~\cite{SharapovaPRA2015,horoshko_bloch_2019,houde_matrix_2024,fabre_modes_2020} (see Sec.~\ref{sec:Partitioning} for the role of the bilinear regime in setting the validity of the partitioning result).

In pulsed SPDC, the overall brightness is set by the parametric gain, but in general, each Schmidt-mode pair has its own gain $G_n$, and the mean photon number generated per pump pulse in the $n$-th pair is $\langle N_n \rangle = \sinh^2 G_n$, so that the detected photon number per pulse is the sum over all occupied modes,
\begin{equation}
    \langle N_{\rm pulse} \rangle = \sum_{n=1}^{\infty} \sinh^2(G_n).
    \label{eq:N_BSV}
\end{equation}
The gain scales with interaction strength and pump intensity as
\begin{equation}
    G_n\propto\chi^{(2)}L\sqrt{I_\mathrm{P}} \, \Gamma_n , 
    \label{eq:G_param}
\end{equation}
where $\chi^{(2)}$ is the second-order susceptibility, $L$ is the interaction length, $I_{\rm P}$ is the pump peak intensity (proportional to the pulse energy for a given pulse duration), and $\Gamma_n$ is the (dimensionless) mode-overlap/phase-matching factor for the $n$-th Schmidt mode. Ultrafast pumping, therefore, makes the regime $\ev*{N_\mathrm{pulse}}\gg1$ readily accessible.

\section{Entanglement in Bright Twin Beams}\label{sec:Partitioning}

The Schmidt-mode framework of Sec.~\ref{sec:Modes} characterizes the mode content of the twin-beam state through $K$, but does not, by itself, address how the bipartite entanglement between signal and idler is distributed across degrees of freedom that are operationally distinct from the experimenter's perspective. Two such categories must be distinguished: \emph{first-quantization} degrees of freedom, indexing the Schmidt mode pairs themselves, and \emph{second-quantization} degrees of freedom (photon number and field quadrature) within each pair of modes.

For bipartite entangled pure states such as BTBs, one very useful witness of entanglement is the \emph{purity} $\mathcal{P}$~\cite{Nielsen2009_QCQI}, or equivalently the \emph{linear entropy} $S_\mathrm{lin}=1-\mathcal{P}$~\cite{Bengtsson2006_GeometryQS}[p.~56] of the reduced single-arm states. For the reduced signal state for instance, they are
\begin{equation}
    \mathcal{P}=\Tr\qty(\hat\rho_\mathrm{s}^2);\qquad S_\mathrm{lin}=1-\Tr\qty(\hat\rho_\mathrm{s}^2),
\end{equation}
where $\hat{\rho}_\mathrm{s}=\Tr_\mathrm{i}\qty(\hat\rho)$ is the partial trace over the idler arm.
Under conditions met by the source demonstrated below, the linear entropy of the reduced single-arm state admits a clean, unique decomposition between these two sectors, and that the Schmidt number $K$ controls the allocation between them.

The argument rests on two structural features of the high-gain parametric process. First, the parametric Hamiltonian is bilinear in the field operators, so its evolution admits a Bloch-Messiah decomposition into independently squeezed Schmidt-mode pairs~\cite{horoshko_bloch_2019,houde_matrix_2024,fabre_modes_2020}. The Schmidt mode \emph{functions} are then gain-independent, and only the per-mode populations $\langle N_n\rangle = \sinh^2 G_n$ evolve with pump power. This separation between fixed mode functions and gain-dependent populations holds provided the experiment remains in the bilinear regime, in which cascaded $\chi^{(2)}$ and other higher-order nonlinearities are kept weak by appropriate choice of crystal length and pump fluence. Second, in the bright limit $\langle N_n\rangle\gg1$ for every occupied mode, each per-mode reduced state is essentially maximally mixed and the linear entropy of the full reduced single-arm state saturates toward unity. Third, the bilinear structure ensures that the purity itself factorizes across the Schmidt basis.
Combining these features, as derived in Supplementary Material Sec. S5, the saturated linear entropy of the reduced single-arm state in the bright few-mode limit decomposes uniquely as
\begin{equation}
    S_\mathrm{lin} \;\simeq\; \underbrace{\frac{K-1}{K}}_{S^\mathrm{eff}_\mathrm{mod}} \;+\; \underbrace{\frac{1}{K}}_{S^\mathrm{eff}_\mathrm{occ}}\,,
    \label{eq:partitioning}
\end{equation}
where $S^\mathrm{eff}_\mathrm{mod}$ is the contribution carried by the first-quantization (mode-index) degrees of freedom, distributed across $K$ orthogonal Schmidt-mode pairs, and $S^\mathrm{eff}_\mathrm{occ}$ is the contribution carried by the second-quantization (photon-number and quadrature) degrees of freedom \emph{within} each pair. The total $S_\mathrm{lin}\simeq1$ is the saturated reference value reached in the bright limit. $K$ controls its allocation between sectors.

The available pump power sets the total resource budget for the bipartite entanglement, but its accessibility under a given detection scheme is determined by which sector carries it. The occupational sector $S^\mathrm{eff}_\mathrm{occ}=1/K$ is the resource exploited by quadrature homodyne, photon-number measurement, and Fock-state heralding (see Supplementary Material Secs. S3, S4): broadband detection projects onto the second-quantization observables of one mode pair without need for mode sorting. Concentrating the pump into a single Schmidt-mode pair ($K\to1$) maximizes the squeezing and redistributes the entanglement budget away from $S^\mathrm{eff}_\mathrm{mod}$ and into $S^\mathrm{eff}_\mathrm{occ}$. The former is the operational resource for multiplexed quantum-information protocols. Each occupied mode pair acts as an independent entangled channel, and the joint state across pairs supports parallel distribution of quantum information across classical degrees of freedom such as frequency, time, or transverse profile. In this latter regime, broadband detection averages over the $K$ independently squeezed pairs and dilutes the observable EPR correlations by a factor of $K$ (see Supplementary Material Sec. S4); accessing the full per-pair entanglement requires mode-selective detection. Single-mode operation is therefore the optimum for direct broadband detection, while controlled multimode operation with a known and stable Schmidt basis is the optimum for multiplexed protocols. It is worth emphasizing the inference hierarchy implicit in this framework. The directly measured quantities in this work are the mean signal photon number $\langle N_{\rm S}\rangle$, the photon-number statistics encoded in $g^{(2)}(0)$, and the spectral covariance of the signal arm. From these we infer an effective Schmidt number $K$, valid under the bilinear high-gain assumption stated above. The decomposition of the saturated reduced-state linear entropy into modal and occupational sectors is then a consequence of $K$, not an independent measurement: for the source demonstrated below, the measured $K\simeq1$ implies that the saturated entropy resides almost entirely in the occupational sector, in the sense made precise by~\eqref{eq:partitioning}. The remainder of this work demonstrates a source that can be deterministically operated in either regime, with $K$ tuned by the pump pulse duration.

\section{Experimental Setup}

\begin{figure*}[t] 
    \centering
    \includegraphics[width=\textwidth]{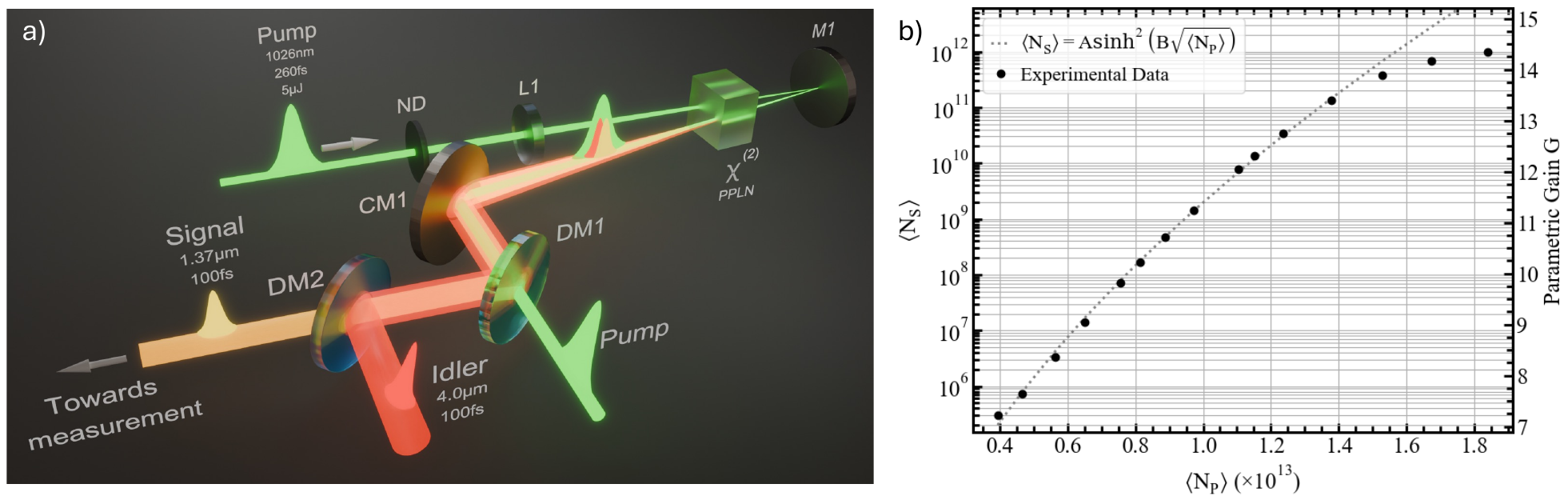}
    \caption{\justifying Generation of ultrafast bright entangled twin-beams. a) Experimental setup. A pump beam (\SI{1026}{\nm}, \SI{260}{\fs}, \SI{5}{\uJ}) is attenuated with a neutral density filter (ND1) and focused into a periodically-poled Lithium Niobate (PPLN) crystal of 2-mm length with a \SI{27.91}{\um} poling period, using a lens (L1) with \SI{250}{\mm} focal length. A mirror (M1) reflects the beam back into the crystal for additional nonlinear interaction. The PPLN is positioned so that the pump's focal plane is at the second pass. The beam is collimated by a concave mirror (CM1) with a focal length of \SI{150}{\mm}. The signal twin (\SI{1.37}{\um}, \SI{100}{\fs}), idler twin (\SI{4.0}{\um}, \SI{100}{\fs}) and pump are separated using a pair of dichroic mirrors (DM1 and DM2). The signal twin is then sent towards measurement devices, while the idler is not studied in this work. b) Average number of signal photons generated per pulse as a function of the average number of pump photons. The dashed line is a fit of the expected number of generated signal photons, showing signs of saturation for $\ev{N_{\rm P}}> \num{1.4e13}$.} 
    \label{fig:fig1}
\end{figure*}

Figure \ref{fig:fig1}a shows the experimental setup used to generate ultrafast bright entangled twin beams via type-0 SPDC in a PPLN crystal. The pump is provided by a Yb:KGW (Light Conversion Pharos) laser (\SI{1026}{\nm} central wavelength, \SI{260}{\fs} pulse duration, \SI{5}{\uJ} pulse energy, up to \SI{1}{\MHz} repetition rate) and is attenuated using a reflective neutral density filter (ND1) before being focused into the PPLN crystal using a \SI{250}{\mm} focal-length lens (L1). The PPLN poling period is \SI{27.91}{\um} and is chosen to produce strongly non-degenerate output with the signal beam centered at \SI{1.37}{\um} and the idler beam at \SI{4.0}{\um} (both $\sim\SI{100}{\fs}$ duration). The available \SI{1}{\MHz} repetition rate represents a significant advancement over previous BSV experiments conducted at \SI{1}{\kHz}~\cite{rasputnyi_hhg_2024,eventzur_attosecond_2025,heimerl_strongfield_2025}, enabling substantially faster acquisition of statistics.

To increase the effective nonlinear interaction and reach higher parametric gain, we employ a folded two-crystal geometry~\cite{perez_2014}: a mirror (M1) retroreflects the pump for a second pass through the $\chi^{(2)}$ crystal. In addition to increasing the parametric gain, such a two-pass configuration was shown to reduce the number of spatial modes by preventing the amplification of strongly diffracting higher-order modes~\cite{perez_2014, cusson_ultrafast_2024}. After the crystal, the co-propagating pump, signal, and idler are collimated by a concave mirror (CM1) with a \SI{125}{\mm} focal length. A pair of dichroic mirrors (DM1 and DM2) then separates the pump from the generated twin beams and splits the signal and idler arms. In the measurements reported here, the signal twin is routed to the diagnostics described below, whereas the idler arm is not analyzed. 

We note that for the chosen wavelength combination, the group-velocity mismatch (GVM) between the pump and signal has the opposite sign of the GVM between the pump and idler; the implications of this for temporal mode selection are discussed in Sec.~\ref{sec:results}.\ref{sec:gdd}.

\section{Results}\label{sec:results}

\subsection{Brightness and Gain Characterization}\label{sec:brightness}

We first characterize the source brightness as a function of pump pulse energy. Figure~\ref{fig:fig1}b shows the measured mean number of signal photons per pulse, $\ev{N_{\rm S}}$, versus the mean number of pump photons per pulse, $\ev{N_{\rm P}}$. The pump level is tuned continuously using the variable attenuation (ND1), and $\ev{N_{\rm P}}$ is obtained from the mean pump pulse energy via $\ev{N_{\rm P}}=\ev{E_{\rm P}}/\hbar\omega_{\rm P}$. The number of photons for a bright twin-beams state depends on the pump intensity, as given by Eqs.~(\ref{eq:N_BSV}-\ref{eq:G_param}). When recording only one arm on a single spatio-temporal mode twin-beam system, one can thus expect a scaling for the mean number of signal photons of the form $\ev{N_{\rm{S}}}=\sinh^2(G)$ where $G\propto\sqrt{\ev{N_{\rm{P}}}}$. Figure \ref{fig:fig1}b highlights a near-perfect match between the experimental data (black dots) and the theoretical fit (dotted line) up to $\ev{N_{\rm{P}}} \simeq\num{1.4e13}$. Above that threshold, the departure from the expected behavior is attributed to the onset of parasitic effects (e.g. pump depletion~\cite{florez_pump_2020}, saturation effects~\cite{spasibko_multiphoton_2017}, or other experimental limitations), and thus delineates the operating range used for the mode-purity analysis below.

The gain in Fig.~\ref{fig:fig1}b ranges from $G\simeq7.6$ at $\ev{N_{\rm P}}=\num{4e12}$ to $G\simeq14.5$ at $\ev{N_{\rm P}}=\num{1.6e13}$. The onset of saturation, observed around $\ev{N_{\rm P}}\simeq\num{1.4e13}$ pump photons/pulse, corresponds to $\ev{N_{\rm S}}\simeq\num{e11}$ signal photons/pulse and $G\simeq 13.5$ under the single-mode estimate. We emphasize that this conversion is provided as a convenient single-parameter scale: in the general multimode case, the detected brightness is given by \eqref{eq:N_BSV}.
For a single-mode operation, such as the one reported in this work (see Section \ref{sec:results}.\ref{sec:g2}), this conversion is exact.

Overall, Fig.~\ref{fig:fig1}b establishes the operating range from the low-brightness regime to strongly amplified emission and provides the reference brightness (and effective gain) scale used to compare mode purity across several orders of magnitude in Secs.~\ref{sec:results}.\ref{sec:g2} and \ref{sec:results}.\ref{sec:cov}.

\subsection{Mode Purity From Photon-Number Statistics}\label{sec:g2}

A direct and experimentally robust probe of spatio-temporal mode structure in bright squeezed vacuum is provided by the second-order intensity correlation function $g^{(2)}(\tau)$~\cite{Glauber1963,chekhova_bright_2015} at zero time delay $\tau = 0$. The correlation function $g^{(2)}(0) = \langle N(N-1) \rangle / \langle N \rangle^2$ quantifies photon-number fluctuations, and in the bright regime ($\langle N \rangle \gg 1$) it is well approximated by $\langle N^2 \rangle / \langle N \rangle^2$.

\begin{figure*}[!t]
    \centering
    \includegraphics[width=\textwidth]{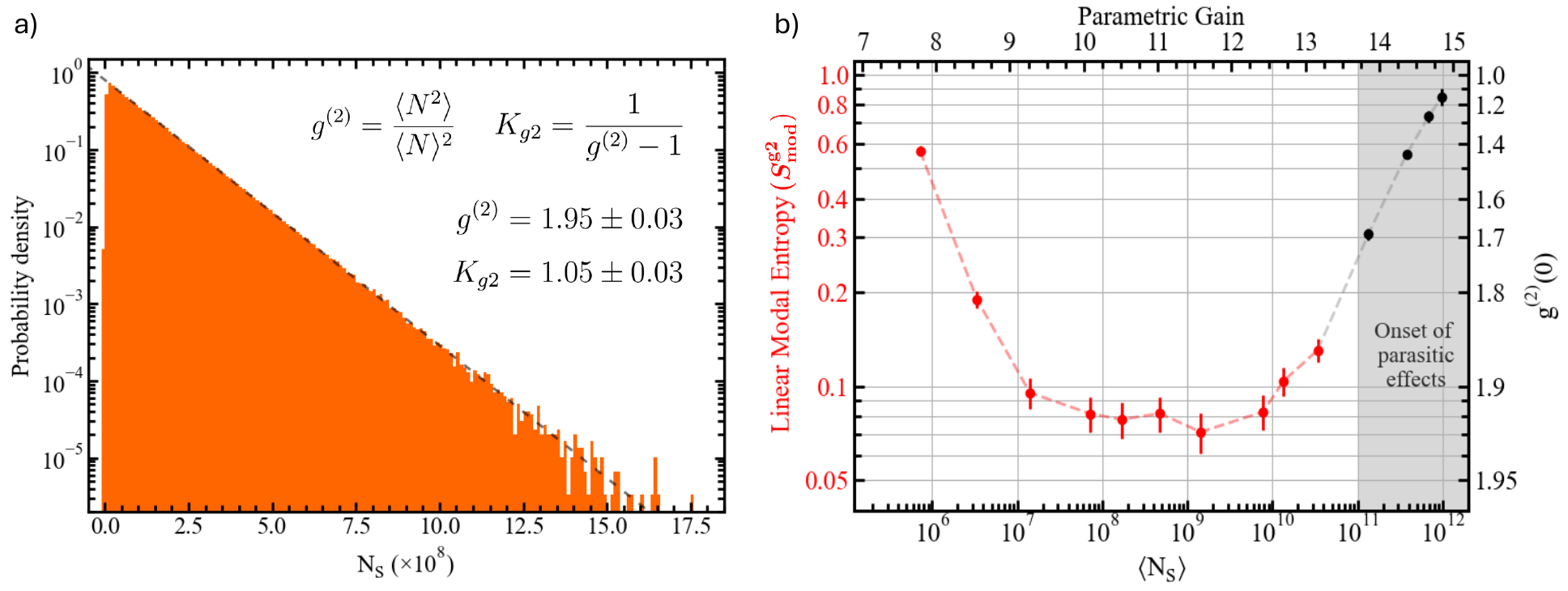}
    \caption{\justifying a) Signal photon number probability density histogram for 2 981 607 recorded events. An average of $\langle N_{\rm S} \rangle= 1.296 \times 10^8$ photons per pulse with a standard deviation of $\sigma_{N_{\rm S}}=1.265 \times 10^8$ photons was measured, giving a parametric gain of $G \simeq 10.3$, Such values give a $g^{(2)}(0)$ value of $1.95 \pm 0.03$, close to the expected $g^{(2)}(0)=2$ for a single-mode thermal state. From this value, a number of spatio-temporal mode of $K_{g2}=1.05 \pm0.03$ was calculated. The dashed line is a guide to the eye showing a perfect thermal (negative-exponential) statistics.
    b) Second-order correlation function $g^{(2)}(0)$ (right axis, black) and the corresponding linear modal entropy $S^{g2}_\mathrm{mod} = (K_{g2}-1)/K_{g2}$ (left axis, red) as a function of mean signal photon number per pulse. Near single-mode operation ($S^{g2}_\mathrm{mod} < 0.1$) is maintained over three orders of magnitude in brightness. The shaded region marks pump levels at which the single-mode sinh² scaling of Fig.~\ref{fig:fig1}b breaks down. The mapping $g^{(2)}(0) \to S^{g2}_\mathrm{mod}$ does not apply in this regime, and the $S^{g2}_\mathrm{mod}$ axis is therefore omitted within the shaded region. The modest rise of $S^{g2}_\mathrm{mod}$ as the shaded boundary is approached from below is attributed to tail-flattening of $P(N)$ by parasitic $\chi^{(2)}$ effects (e.g., SHG of pump or signal) rather than to modal repopulation (see text).}
    \label{fig:fig2}
\end{figure*}

For a single thermal mode, $g^{(2)}(0)=2$; when the detected field is a sum of $K$ statistically independent thermal modes, the bunching is diluted to
\begin{equation}
  g^{(2)}(0) = 1+\frac{1}{K},
  \label{eq:g2_K}
\end{equation}
from which an effective mode number $K_{g2}={1}/({g^{(2)}(0)-1})$ is extracted. It can be interpreted as an inverse-participation ratio of the detected Schmidt-mode distribution. We note that Eq.~(\ref{eq:g2_K}) assumes thermal (Bose--Einstein) statistics within each Schmidt mode, which is the expected distribution for individual spectral-temporal modes of spontaneous PDC at any gain~\cite{Tapster98JMO}.

To evaluate $g^{(2)}(0)$ in the bright regime, we record shot-to-shot signal photon numbers over an ensemble of 2.98 million pulses using a switchable-gain InGaAs photodiode (Thorlabs PDA10CS). The electrical signal from the photodiode was analyzed using a boxcar integrator (Zurich Instruments UHFLI) from which we construct the photon-number histogram $P(N)$ shown in Fig.~\ref{fig:fig2}a. The data is taken at $\ev{N_{\rm P}}= \num{8e12}$ photons, corresponding to an unsaturated gain of $G \sim 10.4$ and an output of $\ev{N_{\rm S}} = \num{1.3e8}$ signal photons per pulse. The measured data are well described by the thermal (negative-exponential) distribution expected for a single mode, for which a guide to the eye is given as the black dashed line in Fig.~\ref{fig:fig2}a. From the same ensemble, we obtain $g^{(2)}(0) = 1.95 \pm 0.03$, corresponding to a linear modal entropy $S^{g2}_\mathrm{mod} = (K_{g2} - 1)/K_{g2} = 0.05 \pm 0.03$, placing approximately 95\% of the saturated bipartite entanglement of the source in the occupational sector (Sec.~\ref{sec:Partitioning}). We note that $g^{(2)}(0)$ of the full beam probes the total number of occupied spatio-temporal modes, including spatial degrees of freedom, whereas the spectral-covariance analysis of Sec.~\ref{sec:results}.\ref{sec:cov} is sensitive only to temporal (frequency) modes due to the spatial filtering imposed by the spectrometer entrance slit.

Figure~\ref{fig:fig2}b shows $S^{g2}_\mathrm{mod}$ extracted as a function of mean photon number per pulse, demonstrating $S^{g2}_\mathrm{mod}<0.1$ over nearly three orders of magnitude in $\langle N_{\rm{S}}\rangle$, forming a ``window'' of optimal single-mode operation.
This trend is consistent with gain selectivity in high-gain SPDC, where increasing amplification preferentially enhances the dominant Schmidt mode(s) and redistributes weight away from weaker modes, thereby reducing the effective mode number toward unity~\cite{SharapovaPRA2015}.
At the highest pump levels, the source itself departs from the single-mode sinh² scaling of $\ev{N_{\rm S}}$ vs. $\ev{N_{\rm P}}$ identified in Fig.~\ref{fig:fig1}b. The shaded region of Fig.~\ref{fig:fig2}b marks this regime: above $\ev{N_{\rm S}} \simeq 10^{11}$ photons per pulse, the simple mapping between $g^{(2)}(0)$ and $K_{g2}$ that underlies the linear entropy inference is no longer applicable, and we do not extract $S^\mathrm{g2}_\mathrm{mod}$ values for these points (the left axis, in red, therefore only applies outside the shaded region). As $\ev{N_{\rm S}}$ approaches the shaded boundary from below, a modest residual rise of $S^\mathrm{g2}_\mathrm{mod}$ is observed in the unshaded data. We attribute this pre-threshold rise not to genuine modal repopulation but to parasitic nonlinear processes that preferentially deplete the strongest pulses, such as cascaded $\chi^{(2)}$ effects (for example, SHG of pump or signal), which flatten the high-photon-number tail of $P(N)$ and reduce $g^{(2)}(0)$ below its single-mode-thermal value~\cite{spasibko_multiphoton_2017}. Because the mean $\ev{N_{\rm S}}$ is insensitive to such tail-flattening, this mechanism is operative throughout the range over which the sinh² fit of Fig.~\ref{fig:fig1}b remains well-satisfied, and is therefore the parsimonious explanation for the unshaded portion of the rise in apparent $S^\mathrm{g2}_\mathrm{mod}$. Overall, these observations indicate that pump strength can be used as a practical knob to approach single-mode operation within a clearly bounded window, beyond which higher-order intensity statistics rather than modal structure govern the apparent value of $g^{(2)}(0)$.

\subsection{Mode Purity From Spectral Covariance}\label{sec:cov}

PDC sources are known to contain strong frequency correlations, which can be modified by the multi-mode nature of the light. The effective number of modes is imprinted in the spectral covariance of the source. In general, and for approximately Gaussian joint spectra, an effective number of spatio-temporal modes can be inferred from the ratio of widths of the conditional and marginal distributions of the joint spectral amplitude which is closely related to the Schmidt number~\cite{Fedorov2009JPB, Law2004PRL}, with the ratio of the marginal to conditional spectral widths corresponding to the Fedorov ratio~\cite{Fedorov2009JPB}.

\begin{figure}[ht!]
    \centering
    \includegraphics[width=0.7\linewidth]{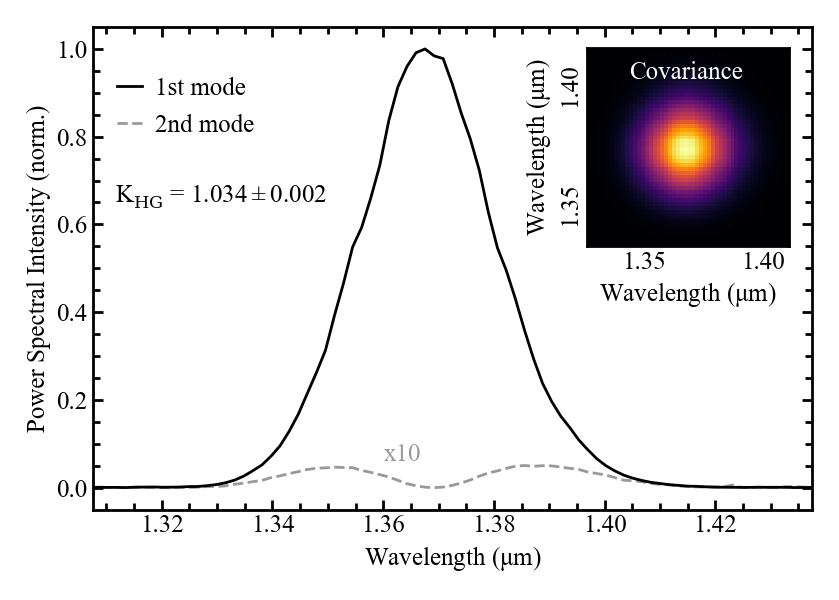}
    \caption{\justifying Temporal mode extraction by singular value decomposition on the reduced spectral density matrix of the signal at high-gain. First and second ($\times 10$) intensity Schmidt modes of the signal. The decomposition yields the Schmidt number $K_\mathrm{HG} = 1.034 \pm 0.002$. Inset: reduced spectral density matrix of the signal, displaying near-perfect symmetry, as expected for a single-spatio-temporal mode light.}
    \label{fig:fig3}
\end{figure}

The reduced spectral density matrix $G^{(1)}$ can be reconstructed by calculating the spectral covariance of one of the twins, in this case the signal measured using an Ocean Optics NIRQuest spectrometer, which is shown in the inset of Figure \ref{fig:fig3}. Using a singular-value decomposition of $G^{(1)}$, the temporal/frequency Schmidt number at high-gain $K_\mathrm{HG}=1.034 \pm0.002$ was calculated (see Supplementary Material Sec. S2.2 for details on the procedure), corresponding to a linear modal entropy $S^\mathrm{HG}_\mathrm{mod} = (K_\mathrm{HG}-1)/K_\mathrm{HG} = 0.034 \pm 0.002$, with approximately 97\% of the saturated bipartite entanglement concentrated in the occupational sector by this measure. The first two Schmidt modes are presented in Figure \ref{fig:fig3}. The uncertainty in the temporal/frequency Schmidt number $K_\mathrm{HG}$ reflects the statistical precision of the decomposition procedure, estimated via bootstrap resampling \cite{efron_bootstrap_1979} over 60 independent subsets of the spectral data, and does not include systematic contributions such as spatial-mode filtering by the spectrometer entrance slit. The two diagnostics agree within their respective uncertainties; the slightly higher $K_\mathrm{g2}$ returned by the $g^{(2)}$ measurement is consistent with its sensitivity to spatial modes, which are filtered out by the spectrometer entrance slit in the $G^{(1)}$ analysis (see Sec.~\ref{sec:results}.\ref{sec:g2}).

We note that near-single-mode operation ($K_\mathrm{HG} \simeq1$) also carries direct implications for the temporal structure of the generated pulses. When the temporal state of the signal is dominated by a single Schmidt mode $\psi_1(t)$ (see \eqref{eq:psiphit}), every shot populates the same spectral profile, differing only in overall amplitude and carrier-envelope phase. Any chirp accumulated during propagation is common to all shots and can be fully compensated by a passive dispersive element to reach the transform limit. The Fourier relationship between frequency and time is then one-to-one: each spectral component maps to a unique temporal feature, enabling unambiguous reconstruction of the temporal field structure from frequency-domain measurements. For $K_\mathrm{HG}>1$, each Schmidt mode is populated with independent random amplitude and phase on every shot. Although all modes acquire a common spectral phase during propagation through shared optics, the random relative phases between modes produce shot-to-shot fluctuations in the composite temporal wavepacket that are equivalent to stochastic variations of group delay, group-delay dispersion, and higher-order dispersion --- fluctuations that no static compensator can remove. A given frequency bin then receives incoherent contributions from multiple modes, rendering the frequency-to-time mapping one-to-many and the temporal structure unrecoverable from spectral data alone. Near-unit spectral purity therefore guarantees both compressibility to the shortest possible pulse duration and a faithful frequency-to-time correspondence, maximizing the peak quantum field amplitude and enabling time-domain quantum-optical measurements that rely on this one-to-one mapping. The source demonstrated here, operating at $K_\mathrm{HG} \simeq 1$ with ultrafast pulse duration, therefore provides direct access to time-domain ultrashort quantum optics with bipartite entanglement --- a regime in which the full temporal structure of the entangled state can be both controlled and resolved.

\subsection{Role of Pump Pulse Duration on Modal Entanglement Entropy} \label{sec:gdd}

The preceding sections establish near-single spatio-temporal mode operation over a broad brightness range under transform-limited pumping. We now show that the mode purity can be deliberately tuned by controlling the \emph{pulse duration} of the pump, without changing its spectral content; within the bilinear-regime partitioning of Sec.~\ref{sec:Partitioning}, this tunability of $K_\mathrm{HG}$ is equivalent to a continuous redistribution of the saturated bipartite entanglement between the occupational and modal sectors via a single experimental parameter.

In the frame co-moving with the pump pulse, parametic process amplifies vacuum fluctuations over a temporal window defined by the pump envelope~\cite{riek_direct_2015,moskalenko_paraxial_2015,Riek2017Nature,BeneaChelmus2019Nature}. The phase-matching bandwidth $\Delta\omega_{\rm PM}$ of the nonlinear medium sets a coherence time $\tau_{\rm c} \sim 1/\Delta\omega_{\rm PM}$ for the amplified fluctuations; the number of orthogonal temporal Schmidt modes that fit within the gain window is then approximately the ratio of the \emph{effective} temporal gain window $\tau_{\rm eff}$ to $\tau_{\rm c}$. Crucially, $\tau_{\rm eff}$ can be substantially shorter than the pump pulse duration $\tau_{\rm p}$ due to two mechanisms. First, the exponential scaling of parametric amplification --- $\langle N(t) \rangle = \sinh^2 G(t)$, where $G(t) \propto \sqrt{I_{\rm P}(t)}$ follows the pump intensity envelope, where it compresses the effective gain window well below $\tau_{\rm p}$. For a Gaussian pump at peak gain $G_0 \gg 1$, $\tau_{\rm eff}$ scales approximately as $\tau_{\rm p}/\sqrt{G_0}$, so that at $G_0 \sim 10$ the effective window can approach $\tau_{\rm c}$ even when $\tau_{\rm p}$ exceeds it by a moderate factor. Second, the strongly non-degenerate geometry provides a complementary constraint through group-velocity mismatch (GVM): for our wavelength combination, the GVM between the pump and signal has the opposite sign of the GVM between the pump and idler, so that the signal walks off ahead of the pump while the idler walks off behind (or vice versa). This limits the temporal window over which all three waves overlap, further restricting the number of modes experiencing significant gain~\cite{cerullo_ultrafast_2003}. In the present source, these two mechanisms together bring $\tau_{\rm eff}$ close to $\tau_{\rm c}$ for the transform-limited \SI{260}{\fs} pump, consistent with the observed $K_\mathrm{HG}\simeq1$. Conversely, stretching the pump via group-delay dispersion (GDD) defeats both mechanisms: the broadened envelope directly increases the temporal gain window, while the reduced peak intensity weakens the exponential gain narrowing that would otherwise compress it. The transition from single-mode to multimode operation with increasing GDD therefore reflects the compounding loss of both mode-filtering effects.

\begin{figure*}[!t]
    \centering
    \includegraphics[width=\textwidth]{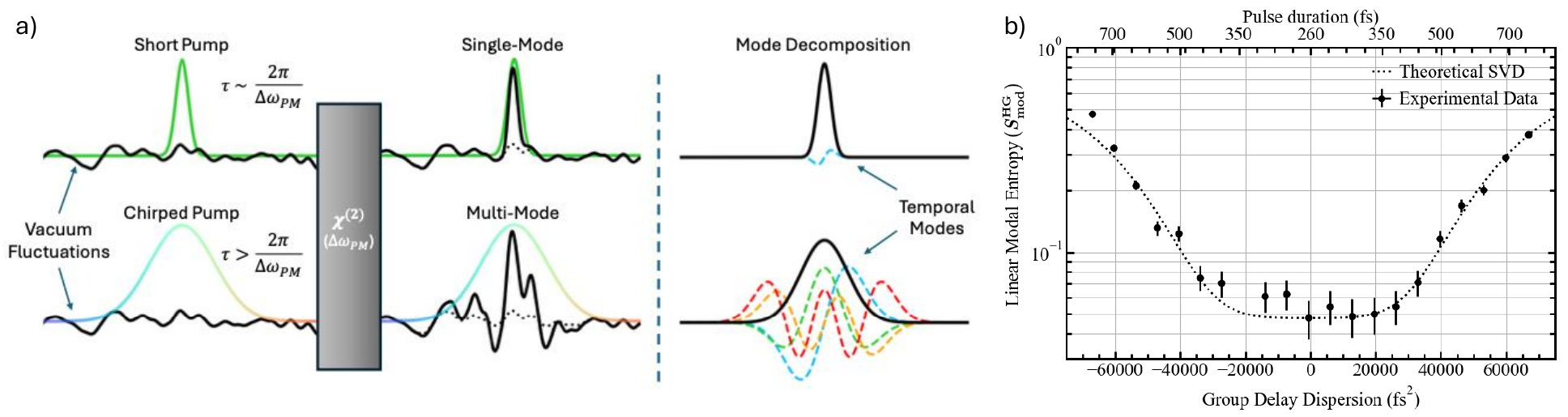}
    \caption{\justifying \textbf{Temporal-mode control by the pump mode} (a) Effect of pump temporal extent on mode purity. Pictorial representation of the vacuum fluctuations available for amplification by SPDC, dictated by the phase matching bandwidth ($\Delta \omega_{PM}$) of the nonlinear medium used. For proper single-mode amplification, the pump duration has to be similar to the inverse of $\Delta \omega_{PM}$. If the pump temporal extent is increased beyond that criterion by the addition of group delay dispersion, multiple temporal modes can be amplified without changing the spectral content of the pump; (b) Measured linear modal entropy $S^\mathrm{HG}_\mathrm{mod} = (K_\mathrm{HG}-1)/K_\mathrm{HG}$ as a function of group delay dispersion (GDD) of the pump, with associated pulse duration. Near-unit purity ($S^\mathrm{HG}_\mathrm{mod}\simeq0$) is maintained only for relatively small added GDD, confirming that single spatio-temporal mode operation is only possible near the transform limit. The methodology for the fit is explained in Supplementary Material Sec. S2. A single fit parameter (minimum entropy) was used to take experimental reality into account, as the theory gives $\min(K_\mathrm{HG})=1$.}
    \label{fig:fig4}
\end{figure*}  

Experimentally, we first match the pump pulse duration to $\tau_{\rm c}$, which is ensured for 260~fs pulses by selecting a PPLN length of 2 mm for our poling period. We then vary the pump GDD over a wide range while keeping the pump spectrum fixed. The pump pulse energy is held constant across the GDD scan; we note that since the pump peak intensity decreases as the pulse is stretched, the parametric gain is also reduced for larger $\abs{{\rm GDD}}$. Because the Schmidt number is itself gain-dependent~\cite{SharapovaPRA2015}, the observed increase in $K$ reflects the combined effect of temporal window broadening and gain reduction. Summarized in Fig.~\ref{fig:fig4}b, the measurements show the extracted linear modal entropy $S^\mathrm{HG}_\mathrm{mod} = (K_\mathrm{HG}-1)/K_\mathrm{HG}$ as a function of applied GDD (bottom axis) and the corresponding pump pulse duration (top axis). The data reveal a clear optimum near the shortest (near-transform-limited) pump duration, when ${\rm GDD}=0$. The data can be compared to the expected linear modal entropy, calculated from the singular value decomposition of the theoretical joint spectral amplitude for a chirped pump (dotted line) (see Supplementary Material, Secs. S1, S2). The observed increase of $K_\mathrm{HG}$ with GDD confirms that pump temporal shaping constitutes a practical parameter for the spatio-temporal mode content in the high-brightness regime, complementing the selectivity effects discussed above. 

Finally, we note that the observed dependence is not strictly symmetric in GDD, as higher-order dispersion (including that of the PPLN crystal itself) and nonlinear phase accumulation can modify the effective phase matching. The dominant trend is governed by the pump duration: single-mode operation is matched to the correlation time set by $\Delta\omega_{\rm PM}$, while deliberately enlarging the pump temporal window increases the number of amplified temporal Schmidt modes and progressively shifts the saturated bipartite entanglement out of the occupational sector and into the modal sector, of interest for multimode continuous-variable protocols.

\section{Discussion and Conclusion}

We have demonstrated an ultrafast bright twin-beam source producing strongly non-degenerate output, signal at \SI{1.37}{\um} and idler at \SI{4.0}{\um}, via type-0 SPDC in a PPLN crystal, operated at up to \SI{1}{\MHz} repetition rate. Two independent diagnostics confirm near-single spatio-temporal mode operation. First, photon-number statistics yield $K_{g2} = 1.05 \pm 0.03$ from the second-order correlation function, characterizing the total number of spatio-temporal modes. In addition, singular-value decomposition of the reduced spectral density matrix of the signal in the high-gain regime gives $K_\mathrm{HG} = 1.034 \pm 0.002$, accounting only for the frequency/time degree of freedom. The corresponding linear entanglement entropies, $S^{g2}_\mathrm{mod} = 0.05 \pm 0.03$ and $S^\mathrm{HG}_\mathrm{mod} = 0.033 \pm 0.002$ respectively, confirm near-unit modal purity maintained over nearly three orders of magnitude in brightness, from $\sim$\num{e7} to $\sim$\num{e10} signal photons per pulse. In this regime, broadband detection is inherently mode-matched, and the full continuous-variable entanglement between signal and idler is accessible without mode sorting.

We have further shown that the pump pulse duration serves as a practical and deterministic control parameter for mode purity: single spatio-temporal mode operation is achieved when the pump duration is matched to the inverse phase-matching bandwidth of the nonlinear medium, while deliberate temporal stretching via added group-delay dispersion drives a controlled transition to multimode emission. This behavior is consistent with the space-time volumetric amplification picture, in which the number of amplified temporal modes is determined by the ratio of the pump temporal window to the phase-matching coherence time. The opposite-sign group-velocity mismatch between the pump--signal and pump--idler pairs and the compression of the temporal gain window act together as a natural temporal mode filter that contributes to the observed near-single-mode operation.

The strong wavelength non-degeneracy of the source naturally separates the mid-infrared idler, which spectrally overlaps with molecular vibrational fingerprints, from the near-infrared signal, which is compatible with efficient, low-noise detection. While the present work focuses on single-arm characterization of the signal twin, direct characterization of the mid-infrared idler and demonstration of twin-beam correlations across the NIR--MIR spectral divide remain as important next steps.

Looking ahead, the combination of ultrafast pulse format, high brightness, MHz repetition rate, and near-unit purity demonstrated here provides a well-defined quantum input state for nonlinear interferometry, quantum-enhanced spectroscopic sensing, and time-domain quantum optics, where the one-to-one frequency-to-time mapping guaranteed by single-mode operation enables direct access to the temporal structure of bipartite entanglement. Single-mode operation is equally enabling for discrete-variable protocols (and their high-brightness extensions~\cite{virally_enhanced_2021}), where photon-number (brightness) conditioning on the near-infrared signal can project the mid-infrared idler onto pure Fock states (band-conditioned states), a route to non-Gaussian quantum state engineering in a spectral region where such states have not yet been demonstrated. Conversely, the same source operated with controlled $K>1$ via pump GDD (Sec.~\ref{sec:results}.\ref{sec:gdd}) provides a deterministic route to multimode bright twin beams with a stable Schmidt basis, of interest for multiplexed continuous-variable protocols in which the modal sector of the entanglement budget becomes the operational resource. More broadly, bright squeezed vacuum has already been shown to drive high-harmonic generation, multiphoton electron emission, and quantum state transfer onto attosecond pulses. The strongly non-degenerate twin-beam source demonstrated here extends these capabilities by combining mid-infrared interaction with near-infrared detection, while preserving the non-classical correlations and temporal mode structure across the full spectral divide.

\section*{Acknowledgments}
The authors acknowledge fruitful conversations with G. Leuchs and N. Quesada.

\subsection*{Author Contributions} 
D. Seletskiy, S. Virally and M. Chekhova conceived the idea.
D. Seletskiy, P. Cusson, G.Demontigny, E. Martel  and A. Rasputnyi designed the experiments.
G. Demontigny, A. Perraton-Elorza, E. Murillo-Zapata, E. Martel and A. Rasputnyi conducted the experiments.
S. Virally, G. Demontigny, D. Seletskiy, A. Rasputniy and M. Chekhova worked on the theory and analyzed the results.
All authors contributed equally to the writing of the manuscript.

\subsection*{Funding}
This research received contribution funding from the Collaborative Science, Technology and Innovation Program of the National Research Council, Canada [agreement QSP-114-1].
This research was supported by the European Union's Horizon Europe Research and Innovation Programme through project MIRAQLS [agreement 101070700],
and by the Fonds de Recherche du Qu\'ebec--Nature et Technologies via Institut Transdisciplinaire d'Information Quantique [DOI 10.69777/340940].
A. Perraton-Elorza received scholarship support from the Natural Sciences and Engineering Research Council of Canada Undergraduate Student Research Award.
E. Murillo-Zapata received scholarship support from the Mitacs Globalink Research Internship program.
M. Chekhova and A. Rasputniy acknowledge funding support from the bilateral Agence Nationale de la Recherche--Deutsche Forschungsgemeinschaft project GENIOUS [DFG project-ID 545591821, ANR-24-CE92-3050050-01].
D. Seletskiy and S. Virally acknowledge funding support from the Natural Sciences and Engineering Research Council of Canada [RGPIN-2025-07169, RGPIN-2026-06186].
D. Seletskiy acknowledges funding support from the Canada Research Chair program [CRC-2021-00102]. 

\subsection*{Competing interests}
The authors declare that there is no conflict of interest regarding the publication of this article.

\subsection*{Data Availability}
Data underlying the results presented in this paper are
not publicly available at this time but may be obtained from the authors upon
reasonable request.

\section*{Supplementary Material -- Materials and Methods}

\renewcommand{\thesection}{S\arabic{section}}
\renewcommand{\theequation}{S\arabic{equation}}
\renewcommand{\thefigure}{S\arabic{figure}}
\renewcommand{\thetable}{S\arabic{table}}
\setcounter{section}{0}
\setcounter{equation}{0}
\setcounter{figure}{0}
\setcounter{table}{0}

\section{SPDC in PPLN with group-delay dispersion}\label{SM:GDD}

We model collinear type-0 spontaneous parametric down-conversion (SPDC) in a periodically poled lithium niobate (PPLN) crystal pumped by a \unit{\fs} laser, including group-delay dispersion (GDD).

\subsection{Material dispersion and phase matching}\label{SM:GDD:Disp}

The extraordinary refractive index of the lithium-niobate crystal as a function of wavelength and temperature is provided by its manufacturer~\cite{Covesion2026_PPLN}:
\begin{equation} 
n^2(\lambda) = a_1+f_T\,b_1 + \frac{a_2+f_T\,b_2}{\lambda^2 - \qty(a_3 + f_T\,b_3)^2} + \frac{a_4+f_T\,b4}{\lambda^2-a_5^2}-a_6\,\lambda^2,
\label{SM:n2}
\end{equation}
with $\lambda$ in \unit{\um}, parameters
$a_1=5.756,a_2=0.0983,a_3=0.202,a_4=189.32,a_5=12.52,a_6=0.0132$, $b_1=\num{2.86e-6},b_2=\num{4.7e-8},b_3=\num{6.113e-8},b_4=\num{1.516e-4}$, and $f_T=T^2-T^2_\mathrm{room}$, where $T$ is the temperature of the crystal and $T_\mathrm{room}$ the room temperature, in \unit{\kelvin}.

This determines the wavenumbers $k(\omega) = n(\lambda)\,\omega/c$.

Energy conservation imposes $\omega_\mathrm{p} = \omega_\mathrm{s} + \omega_\mathrm{i}$.
The quasi-phase-matching condition (momentum conservation) is~\cite{Armstrong1962_PhysRev}
\begin{equation}
\Delta k(\omega_\mathrm{s},\omega_\mathrm{i}) = k_\mathrm{p}(\omega_\mathrm{s}+\omega_\mathrm{i}) - k_\mathrm{s}(\omega_\mathrm{s}) - k_\mathrm{i}(\omega_\mathrm{i}) - \frac{2\pi}{\Lambda},
\label{SM:Dk}
\end{equation}
where $\Lambda$ is the poling period.
The phase-matching function is
\begin{equation}
\Phi(\omega_\mathrm{s},\omega_\mathrm{i}) = \mathrm{sinc}\qty[\frac{\Delta k(\omega_\mathrm{s},\omega_\mathrm{i})\,L}{2}]
\exp\qty[i\frac{\Delta k(\omega_\mathrm{s},\omega_\mathrm{i})\,L}{2}],
\label{SM:Phi}
\end{equation}
where $L$ is the length of the crystal.

\subsection{Pump envelope and GDD}\label{SM:GDD:Pump}

The pump is modeled as a Gaussian pulse. In the frequency domain, using detunings $\Omega = \omega - \omega_0$,
\begin{equation}
\alpha(\Omega_\mathrm{p}) = \exp\!\left[-\frac{\Omega_\mathrm{p}^2}{2\sigma_\mathrm{p}^2}\right]
\exp\!\left(i \frac{\mathrm{GDD}}{2} \Omega_\mathrm{p}^2 \right).
\label{SM:alpha}
\end{equation}

The quadratic phase term arises from group-delay dispersion (GDD)~\cite{RPPhotonics2026_GDD}.
Importantly, since $\Omega_\mathrm{p}=\Omega_\mathrm{s}+\Omega_\mathrm{i}$ it contains a \emph{non-separable} phase term proportional to $\Omega_\mathrm{s} \Omega_\mathrm{i}$, which modifies the modal structure.

\section{Measurements of the modal structure}\label{SM:Modes}

\subsection{Low-gain regime: joint spectral amplitude and first-order correlation function}\label{SM:Modes:LG}

For simplicity, herein we concentrate on the frequency/time degrees of freedom and trace out the transverse spatial degrees of freedom in the first-quantization picture.
A generalization to all degrees of freedom is fairly obvious.
The low-gain biphoton state is~\cite{Brecht2015_PRX}
\begin{equation}
\ket{\Psi}=\iint \dd\omega_\mathrm{s}\dd\omega_\mathrm{i}\,
J(\omega_s,\omega_i)\,
\hat a_\mathrm{s}^\dagger(\omega_\mathrm{s})\hat a_\mathrm{i}^\dagger(\omega_\mathrm{i})\ket{0},
\label{SM:PsiInt}
\end{equation}
where $J(\omega_\mathrm{s},\omega_\mathrm{i})= \alpha(\omega_\mathrm{s}+\omega_\mathrm{i})\,\Phi(\omega_\mathrm{s},\omega_\mathrm{i})$ is the joint spectral amplitude (JSA).

The JSA admits the Schmidt decomposition
\begin{equation}
J(\omega_\mathrm{s},\omega_\mathrm{i})=\sum_n \sqrt{\lambda_n}\,\phi_n(\omega_\mathrm{s})\,\psi_n(\omega_\mathrm{i}),
\label{SM:JSA}
\end{equation}
with orthonormal mode functions $\{\phi_n\}$ and $\{\psi_n\}$, and singular values $\sqrt{\lambda_n}\ge0$, verifying $\sum_n \lambda_n=1$.

We define the broadband creation operators
\begin{equation}
\hat{A}_n^\dagger=\int\dd\omega\,\phi_n(\omega)\hat{a}_\mathrm{s}^\dagger(\omega),
\qquad
\hat{B}_n^\dagger=\int \dd\omega\,\psi_n(\omega)\hat{a}_\mathrm{i}^\dagger(\omega).
\label{SM:A_nB_n}
\end{equation}
Then
\begin{equation}
\ket{\Psi}=\sum_n\sqrt{\lambda_n}\,\hat{A}_n^\dagger\,\hat {B}_n^\dagger\,\ket{0}.
\label{SM:PsiSum}
\end{equation}

Tracing over the idler gives the reduced density operator of the signal,
\begin{equation}
\hat\rho_\mathrm{s}=\Tr_\mathrm{i}\qty(\dyad{\Psi})=\sum_n \lambda_n\,\dyad{\phi_n}.
\label{SM:rho_s}
\end{equation}

The first-order correlation function of the signal is then
\begin{equation}
G_\mathrm{s}^{(1)}(\omega,\omega')=
\Tr\qty[\hat{\rho}_\mathrm{s}\,
\hat{a}_\mathrm{s}^\dagger(\omega)\hat{a}_\mathrm{s}(\omega')]=
\sum_n\lambda_n\,\phi_n(\omega)\,\phi_n^*(\omega').
\end{equation}

Therefore, the eigenfunctions of $G_s^{(1)}$ are the signal Schmidt modes $\phi_n$, with corresponding eigenvalues $\lambda_n$, the squares of the singular values of the JSA.

The effective Schmidt number at low gain is defined as
\begin{equation}
K_\mathrm{LG}=\frac{1}{\sum_n\lambda_n^2}.
\label{SM:K_LG}
\end{equation}

\subsection{High-Gain Regime and Gain Narrowing}\label{SM:Modes:HG}

At high parametric gain, each Schmidt mode evolves independently (Bloch--Messiah decomposition)~\cite{Perez2015_PRA}.
The squeezing parameter for mode $n$ scales as
\begin{equation}
r_n \propto G \sqrt{\lambda_n}.
\label{SM:r_n}
\end{equation}

The mean photon number in each mode is then 
$N_n = \sinh^2(r_n)$, and the normalized modal weights are
$\pi_n = N_n/\sum_m N_m$.

The effective mode number in the bright regime becomes
\begin{equation}
K_{\mathrm{HG}} = \frac{1}{\sum_n \pi_n^2}.
\label{SM:K_HG}
\end{equation}

Because $\sinh^2$ is strongly nonlinear, the largest initial Schmidt eigenvalues dominate, leading to \emph{gain narrowing}:
\begin{equation}
K_{\mathrm{HG}} < K_{\mathrm{LG}}.
\label{SM:GainNarrowing}
\end{equation}

\subsection{Relation to $g^{(2)}(0)$ measurements}\label{SM:Modes:g2}

The effective Schmidt numbers above are obtained purely from spectral correlations.
Given the size of the entrance slit of the spectrum analyzers, these measurements trace out the transverse spatial degrees of freedom.

A different method can be used to estimate the total number of spatio-temporal modes.
Instead of measuring spectra, it consists in measuring intensities in large aperture detectors.
It requires the additional assumption that each mode is populated with thermal statistics.
This assumption is normally valid for non-degenerate SPDC processes when the twin branch is traced out.

For a mixture of independent thermal modes with relative populations $\pi_n$ $(\sum_n\pi_n=1)$, the normalized second-order correlation function at zero delay, $g^{(2)}(0)=\ev{I^2(t)}/\ev{I(t)}^2$ satisfies
\begin{equation}
g^{(2)}(0)=1+\sum_n\pi_n^2,
\label{SM:g2}
\end{equation}
so that an effective Schmidt number can be computed as
\begin{equation}
K_{g2}=\frac{1}{g^{(2)}(0)-1}.
\label{SM:K_g2}
\end{equation}

Since $G^{(1)}$ probes only longitudinal modes and $g^{(2)}$ captures all spatio-temporal modes, we expect that $K_{\mathrm{g2}}>K_\mathrm{LG}$ in the low-gain regime and 
$K_{\mathrm{g2}}>K_\mathrm{HG}$ in the high-gain regime.

\section{Accessibility of quadrature entanglement in multimode SPDC}\label{SM:Quads}

An important aspect of the single mode source described in the main text is that it allows full access to quadrature entanglement without mode sorting, experimentally a complex task.
We show here how multimode sources mix and significantly degrade the quantum correlations between simultaneous quadrature measurements on the twin beams.

In a homodyne measurement, the signal and idler fields are projected onto the temporal modes defined by the local oscillators (LOs) $u_\mathrm{s}(t)$ and $u_\mathrm{i}(t)$, which we consider classical.
Writing the field operators in the Schmidt basis,
\begin{equation}
\hat{a}_\mathrm{s}(t)=\sum_n\phi_n(t)\hat{A}_n,\qquad
\hat{a}_\mathrm{i}(t)=\sum_n\psi_n(t)\hat{B}_n,
\label{SM:a_sa_i}
\end{equation}
the homodyne detectors measure the effective modes
\begin{equation}
\hat{A}=\int\dd t\,u_\mathrm{s}^*(t)\hat{a}_\mathrm{s}(t)\equiv\sum_n \alpha_n\hat{A}_n,\qquad
\hat{B}=\int\dd t\,u_\mathrm{i}^*(t)\hat{a}_\mathrm{i}(t)\equiv\sum_n \beta_n\hat{B}_n,
\label{SM:AB}
\end{equation}
with overlap coefficients
\begin{equation}
\alpha_n=\int\dd t\,u_\mathrm{s}^*(t)\phi_n(t),
\qquad
\beta_n=\int\dd t\,u_\mathrm{i}^*(t)\psi_n(t).
\label{SM:c_nd_n}
\end{equation}

The detected quadratures are therefore linear combinations of independent squeezed modes.
The variance of the difference quadrature can be written as
\begin{equation}
\mathrm{Var}(\hat{X}_\mathrm{s}-\hat{X}_\mathrm{i})=1+\sum_n \qty(\abs{\alpha_n}^2+\abs{\beta_n}^2)\sinh^2 r_n-
2\,\Re\qty[e^{-i(\theta_\mathrm{s}+\theta_\mathrm{i})}
\sum_n \alpha_n \beta_n\,e^{i\varphi_n}\sinh r_n\cosh r_n],
\label{SM:VarFull}
\end{equation}
where $\theta_\mathrm{s}$ and $\theta_\mathrm{i}$ are the LO phases and $\varphi_n$ is the squeezing phase of Schmidt mode $n$.

The sum inside the last term clearly shows that the variance cannot be optimally minimized, unless all squeezing phases are the same, a condition that would be very difficult to fulfill experimentally.

Even in the ideal case where all $\varphi_n$ are equal and the overlap coefficients verify $\alpha_n=\beta_n$ for all $n$, we get
\begin{equation}
\mathrm{Var}(\hat X_\mathrm{s}-\hat X_\mathrm{i})=\sum_n\abs{\alpha_n}^2\,e^{-2r_n},
\label{SM:VarBest}
\end{equation}
which is always worse than the single-mode case where $\mathrm{Var}(\hat{X}_\mathrm{s}-\hat{X}_\mathrm{i})=e^{-2r_1}$.

\section{Effect of the number of modes on photon counting and conditioning on the number of photons}\label{SM:Photons}

In contrast to homodyne measurements, photon counting is inherently multimode. It sums over all Schmidt modes.

The total photon number in each arm is
\begin{equation}
\hat{N}_\mathrm{s}=\sum_n\hat{A}_n^\dagger\hat{A}_n,\qquad
\hat{N}_\mathrm{i}=\sum_n\hat{B}_n^\dagger\hat{B}_n.
\label{SM:N_sN_i}
\end{equation}

Because the SPDC state consists of independent twin-beam pairs, one has perfect pairwise correlations for each Schmidt mode.
Consequently,$\hat{N}_\mathrm{s}=\hat{N}_\mathrm{i}$
in the ideal lossless case, regardless of the number of modes.

However, multimode structure affects higher-order properties.
In particular, conditioning on a photon-number measurement in one arm produces a pure Fock state \emph{only in the single-mode case}, whereas in the multimode case it leads to a statistical mixture over different modal distributions.
This statement can be made explicit by writing the output state in the Schmidt basis as a tensor product of two-mode squeezed states,
\begin{equation}
\ket{\Psi}=\bigotimes_n\sqrt{1-\abs{\mu_n}^2}\sum_{k_n=0}^\infty \mu_n^{k_n}\ket{k_n}_{\mathrm{s},n}\ket{k_n}_{\mathrm{i},n},
\label{SM:TMSSMixed}
\end{equation}
where $\mu_n=e^{i\varphi_n}\tanh r_n$. Expanding over all modal occupations gives
\begin{equation}
\ket{\Psi}=\sum_{\qty{k_n}}C_{\qty{k_n}}\,\ket{\qty{k_n}}_\mathrm{s}\ket{\qty{k_n}}_\mathrm{i},\qquad
C_{\qty{k_n}}=\prod_n\sqrt{1-\abs{\mu_n}^2}\;\mu_n^{k_n}.
\label{SM:PsiMixed}
\end{equation}

A photon-number measurement in the signal arm with outcome $N$ corresponds to the projector
\begin{equation}
\hat{\Pi}_N^{(\mathrm{s})}=\sum_{\qty{k_n}:\,\sum_n k_n=N} \dyad{\qty{k_n}}_\mathrm{s}.
\label{SM:Pi_N}
\end{equation}
The conditional idler state is therefore
\begin{equation}
\hat{\rho}_{\mathrm{i}|N}\propto
\Tr_\mathrm{s}\qty[(\hat{\Pi}_N^{(\mathrm{s})}\otimes\hat{I}_\mathrm{i})\dyad{\Psi}].
\label{SM:rho_i|NDef}
\end{equation}

Using the orthogonality of the multimode Fock basis, one obtains
\begin{equation}
\hat{\rho}_{\mathrm{i}|N}\propto
\sum_{\qty{k_n}:\,\sum_n k_n=N}\abs{C_{\{k_n\}}}^2\,
\dyad{\qty{k_n}}_\mathrm{i}.
\label{SN:rho_i|NRes}
\end{equation}
Hence, conditioning on the total photon number does not select a unique idler Fock state, but rather a mixture over all partitions of $N$ photons among all Schmidt modes.

\section{Effective decomposition of linear entropy into modal and occupational contributions}\label{SM:Entropy}
\subsection{Linear entropy}\label{SM:Entropy:Linear}
In this section, we use the generalized entropy known in the quantum information literature as the linear entropy~\cite{Bengtsson2006_GeometryQS}[p.~56]
\begin{equation}
    S_\mathrm{lin}=1-\Tr\qty(\hat\rho^2).
    \label{SM:S_lin}
\end{equation}

\subsection{Reduced density matrix of the signal}\label{SM:Entropy:Reduced}
As before, we define the modal weights $\pi_n=N_n/\sum_m N_m$, so that
$\sum_n \pi_n = 1$.

Also, the reduced density matrix, once the idler has been traced out, is
\begin{equation}
    \begin{split}
    \hat{\rho}_\mathrm{s}&=\sum_{\qty{k_n}}\prod_n\qty(1-\abs{\mu_n}^2)\abs{\mu_n}^{2k_n}\dyad{\qty{k_n}}_\mathrm{s}\\
    &=\bigotimes_n\qty(1-\abs{\mu_n}^2)\sum_{k_n}\abs{\mu_n}^{2k_n}\dyad{k_n}_{n,\mathrm{s}},
    \end{split}
    \label{SM:rho_s2}
\end{equation}
with $\abs{\mu_n}=\frac{N_n}{N_n+1}$.

\subsection{Hierarchical decomposition of the entropy}\label{SM:Entropy:Hierarchy}
We introduce a hierarchical decomposition of the entropy~\cite{Perotti2020_PRE} by defining modal sectors in the full state, calculating the occupational entropy inside each sector. In the absence of mode sorting, we calculate a weighted occupational entropy. This corresponds to the part of the entropy attributable to the occupational (second quantization) degrees of freedom. The remaining entropy is ascribed to the modal (first quantization) degrees of freedom.

Each signal mode is individually thermal,
\begin{equation}
    \hat\rho_{n,\mathrm{s}}=\qty(1-\abs{\mu_n}^2)\sum_{k_n}\abs{\mu_n}^{2k_n}\dyad{k_n}_\mathrm{s}.
    \label{SM:rho_ns}
\end{equation}

Therefore the linear entropy associated with mode $n$ is
\begin{equation}
    S_n=1-\Tr\qty(\hat\rho_{n,\mathrm{s}}^2)=1-\frac{1}{2N_n+1}=\frac{2N_n}{2N_n+1}.
    \label{SM:S_n}
\end{equation}

The natural weighted occupation entropy of the signal is
\begin{equation}
    S_\mathrm{occ,s}^\mathrm{eff}=\sum_n\pi_n^2\;\frac{2N_n}{2N_n+1}.
    \label{SM:S_2Q}
\end{equation}
Notice that the weights $\pi_n^2$ differ from the usual weights $\pi_n$ for the von Neumann entropy. They stem directly from the fact that $S_\mathrm{lin}$ is a second-order Havrda and Charv\'at (or Tsallis) entropy~\cite{Bengtsson2006_GeometryQS}[p.~56].

Using \eqref{SM:rho_s2}, the full signal linear entropy is
\begin{equation}
    S_\mathrm{s}=1-\Tr\qty(\hat\rho_\mathrm{s}^2)=1-\prod_n\frac{1}{2N_n+1}.
    \label{SM:S_s}
\end{equation}

The effective modal entropy is obtained by subtracting the contribution associated with the occupational degrees of freedom from the total entropy,
\begin{equation}
    S_\mathrm{mod,s}^\mathrm{eff}=S_\mathrm{s}-S_\mathrm{occ,s}^\mathrm{eff}.
    \label{SM:S_1Q}
\end{equation}

\subsection{Bright few-mode limit}\label{SM:Entropy:Bright}
The decomposition is very enlightening in the bright few-mode limit. We consider that a finite number $M$ of modes are occupied, and all occupied modes are bright,
\begin{equation}
    N_n\gg1\quad\forall n\le M;\qquad\qquad N_n\ll1\quad\forall n>M.
    \label{SM:OccupBright}
\end{equation}

Then for all occupied modes, $S_n\simeq1$, and for all unoccupied modes, $S_n\simeq0$. Thus, we have
\begin{equation}
    S_\mathrm{occ,s}^\mathrm{eff}
    \simeq\sum_{n=1}^{M}\pi_n^2\simeq\sum_n\pi_n^2=\frac{1}{K}.
    \label{SM:S_2QBright}
\end{equation}

Also
\begin{equation}
S_\mathrm{s}\simeq1-\prod_{n=1}^M\frac{1}{2N_n}=1-\frac{1}{2^M\prod_{n=1}^M N_n}\simeq1.
\label{SM:S_sBright}
\end{equation}

Then, one finds
\begin{equation}
    S_\mathrm{mod,s}^\mathrm{eff}\simeq1-\frac{1}{K};\qquad\qquad
    S_\mathrm{occ,s}^\mathrm{eff}\simeq\frac{1}{K}.
    \label{SM:S_1QBright}
\end{equation}

Thus, in the bright few-mode regime and in the absence of mode sorting, the linear entropy separates approximately into a modal component $1-1/K$ and an occupational component $1/K$.

The modal component $1-1/K$ is exactly the linear form of the classical entropy
\begin{equation}
    H_\mathrm{lin}\qty(\qty{\pi_n})=1-\sum_n\pi_n^2,
    \label{SM:Shannon}
\end{equation}
so that we have a natural ``sectorial'', or ``coarse-grained'' decomposition
\begin{equation}
    S_\mathrm{lin}=H_\mathrm{lin}\qty(\qty{\pi_n})+
    \sum_n\pi_n^2S_{\mathrm{lin},n},
\end{equation}
where the second part of the sum on the RHS is the weighted average of the entropy for all sectors (the intra-sector entropy) and the first term is the inter-sector entropy.

Hence, our hierarchical entropy decomposition is formally identifiable to the canonical joint entropy~\cite{Nielsen2009_QCQI}[p.~513], as if the full Fock space was made of a direct sum over modal subspaces, although it is not. This is a general idea behind hierarchical models of entropy decomposition, and it fully justifies our interpretation of the decomposition for bright TMSS.

Note that $S_\mathrm{lin}=1$ is the maximal value of the linear entropy (obtained for a Hilbert space with infinite dimension, well approximated by bright TMSS). Hence, the total linear entropy $\sim1$ indicates maximum mixedness in the reduced signal state, and thus maximal entanglement with the idler. In our context of bright BTB without mode sorting, single-modedness allows all this entanglement to be carried by the occupational degrees of freedom. In the opposite case, a highly multimode state would carry almost all the entanglement in the modal degrees of freedom instead.

\printbibliography

\end{document}